\documentclass{aa}

\newcommand{\pri}    {${\rlap.}^{\prime \prime}$}

\newcommand{\ltsima} {$\; \buildrel < \over \sim \;$}
\newcommand{\simlt}  {\lower.5ex\hbox{\ltsima}}            
\newcommand{\gtsima} {$\; \buildrel > \over \sim \;$}
\newcommand{\simgt}  {\lower.5ex\hbox{\gtsima}}            

\usepackage{graphicx}
\usepackage{txfonts}
\begin{document}

   \title{Multi-colour optical photometry of \object{V404 Cygni} in outburst}

   \subtitle{}

   \author{Josep Mart\'{\i}
          \inst{1}
          \and
          Pedro L. Luque-Escamilla\inst{2}
          \and
           Mar\'{\i}a T. Garc\'{\i}a-Hern\'andez\inst{1}
          }

   \institute{Departamento de F\'{\i}sica, Escuela Polit\'ecnica Superior de Ja\'en, Universidad de Ja\'en, Campus Las Lagunillas s/n, A3, 23071 Ja\'en, Spain\\
   \email{jmarti@ujaen.es,tgarcia@ujaen.es}
   \and
   Departamento de Ingenier\'{\i}a Mec\'anica y Minera, Escuela Polit\'ecnica Superior de Ja\'en, Universidad de Ja\'en, Campus Las Lagunillas s/n, A3, 23071 Ja\'en, Spain\\
   \email{peter@ujaen.es}
               }

   \date{Received XXXX, 2015; accepted XXXX,  XXXX}

 
  \abstract
   {This observational paper has been prepared in the context of the large multi-wavelength effort by many observers
   with the aim of following up the transient flaring event of \object{V404 Cygni} that took place for several weeks in 2015 June.}
   {Our main original aim was to contribute to the study of this transient source by acquiring  broad-band photometric observations during its most active flaring phases.
   Nevertheless, after a detailed analysis of the data, several interesting results were obtained that encouraged a dedicated publication.}
   {The methodology used was based on broad-band differential CCD photometry. 
   This outburst of \object{V404 Cygni} rendered the source a very bright target easily within reach of small educational telescopes.
   Therefore,  the 41 cm telescope available at the Astronomical
   Observatory of the University of Ja\'en was used in this work.
   }
   {We detected variability at different time scales,  both in amplitude and colour. 
   Individual optical flares appear every half hour on average
   during our 3 h long observation, although large-amplitude ($\sim 1$ mag) variations are also observed to occur 
   on intervals as short
   as 10 minutes. Also, colour variations appear to be highly correlated in a 
   colour-colour diagram. 
   Another remarkable finding is the detection of time lag, from about one to a fraction of a minute between light curves in different filters ($VR_cI_c$).
   }
   { The observed behaviour is tentatively interpreted in an scenario based on  the ejection of non-thermal emitting, relativistic plasmons,  with their synchrotron spectra extending
   up to optical wavelengths.
   This would render some of the 
   \object{V404 Cygni} flares very similar to those of the well-know microquasar \object{GRS 1915+105}}


   \maketitle
%

\section{Introduction}
 
\object{V404 Cygni}  also known as GS 2023+338 is a low-mass X-ray binary (LMXB)  originally 
discovered during its 1989 outburst by the Japanese {\it GINGA}
satellite \citep{1989Natur.342..518K}. Intensive photometric and spectroscopic follow-up of this historical event
yielded an orbital period of about 6.5 d and remarkably concluded with the presence of a massive stellar black hole in the system
\citep{1992Natur.355..614C,1992ApJ...401L..97W}. The companion star of \object{V404 Cygni} has been proposed to be  an object of a late-type
K0($\pm$1) III-V spectral type
 \citep{1993MNRAS.265..834C}.
Radio detections using Very Long Baseline Interferometry (VLBI)  enabled an accurate parallax measurement that places \object{V404 Cygni}
at a distance of $2.39 \pm 0.14$ kpc \citep{2009ApJ...706L.230M}. These VLBI observations also constrained the size of any quiescent jets to less than 1.4 AU.
The quiescent X-ray spectrum has a power-law photon index $\Gamma \simeq 2.0$ seen through a 
a total column density of $N_H = (1.0 \pm 0.1) \times 10^{22}$ cm$^{-2}$ (see e.g. \citet{2014MNRAS.441.3656R} and references therein).
The unabsorbed 0.3-10 keV  luminosity approaches several times $10^{32}$ erg s$^{-1}$, thus making \object{V404 Cygni} the brightest black hole LMXB in quiescence.

On 16 June 2015, \object{V404 Cygni} was reported to be in outburst by the Swift Burst Alert Telescope  \citep{GCN_circ}, and soon confirmed by the
Monitor of All-sky X-ray Image on board the International Space Station \citep{2015ATel.7646....1N}.
After these early warnings, an intensive observational effort was deployed by many observers that was quickly reflected in an intense
flow of tens of related electronic telegrams. Soon, it became evident that this was an extraordinary outburst event after decades of quiescence.
Optical, infrared, radio, X-ray and gamma-ray telescopes have so far collected an impressive amount of data that will emerge in the scientific literature
in the upcoming months. But the extraordinary thing was that even small optical telescopes could join this endeavor since \object{V404 Cygni}
became a bright and highly variable source at visible wavelengths \citep{2015ATel.7677....1H, 2015ATel.7681....1H, 2015ATel.7688....1W, 2015ATel.7721....1S}.

In this work, we present the optical data acquired using the educational astronomical facilities available 
at the University of Ja\'en (UJA). Our aim is to contribute to the wealth of public astronomical data about this transient event while, at the same time, attempting
to better constrain the nature of this phenomenon.

\section{Observations}

The observations were carried out on  26 June 2015 from the UJA Astronomical Observatory. The observatory is located
in an urban area inside the Campus of Las Lagunillas, and hosts an automated 41 cm Schmidt-Cassgrain with f/8 focal ratio. 
The UJA telescope, heareafter UJT, operates using a ST10-XME commercial CCD camera with $2184 \times 1472$ pixels of 6.8 $\mu$m size. The pixel scale is 0\pri 42 pixel$^{-1}$.
The camera is  equipped with a wheel of $UBVR_cI_c$ Johnson-Cousins filters \citep{1953ApJ...117..313J,  1974MNRAS.166..711C, 1974MNSSA..33..149C} manufactured 
according to the Bessell filter prescription \citep{1979PASP...91..589B}. The seeing average was typically of about $2^{\prime\prime}$, and therefore $2\times2$ binning was used.
 
\begin{figure}
   \centering
   \includegraphics[angle=0,width=9.3cm]{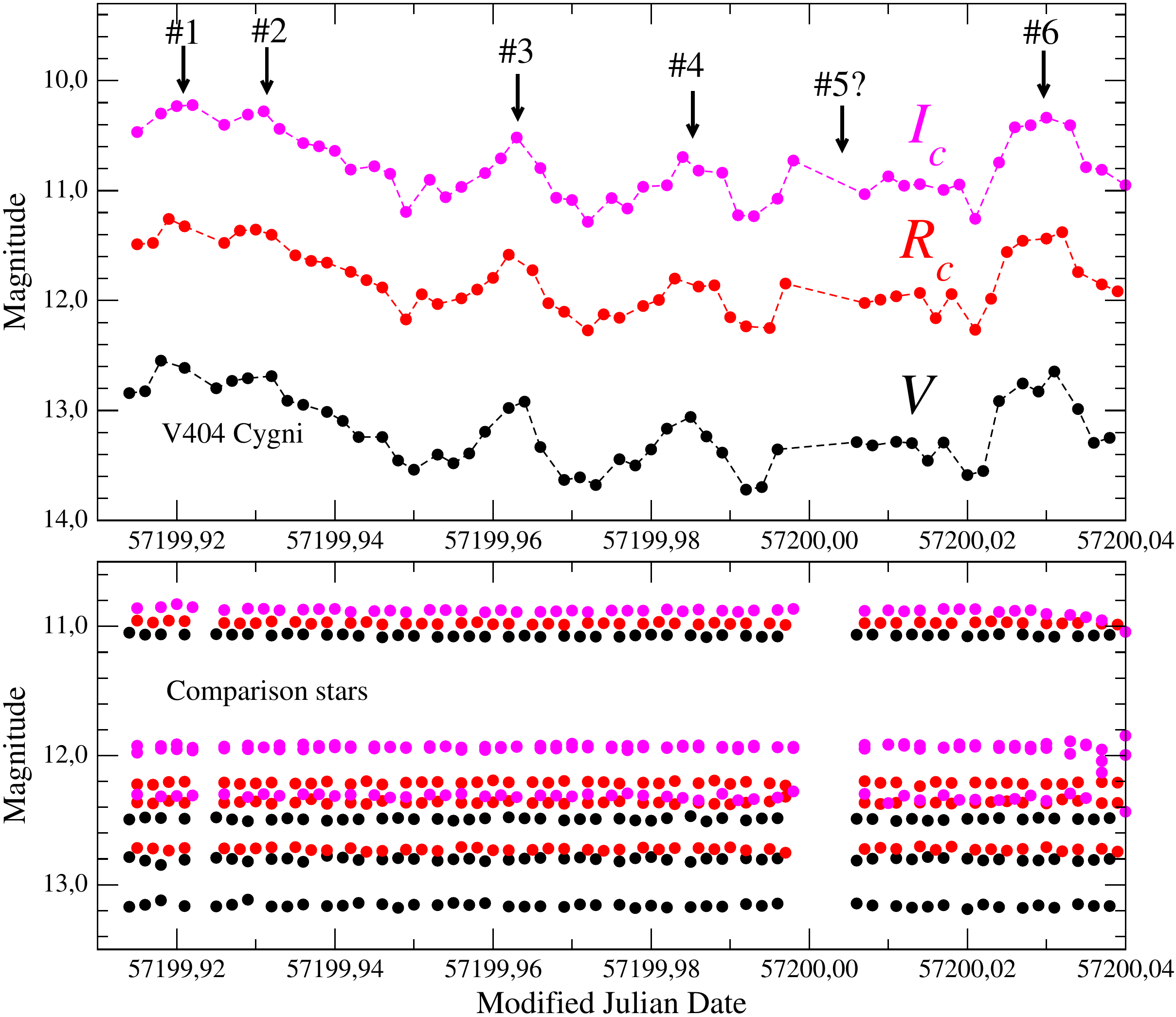}
      \caption{{\bf Top.} Light curves of \object{V404 Cygni} in outburst as observed with the UJT on 26 June 2015  in the $V$, $R$ and $I$ bands.
  Flares are labelled for clarity using vertical arrows.
  {\bf Bottom.} Behaviour of the four comparison stars, plotted at the same scale, which remained constant in brightness within 0.01-0.02 mag.   
    }
         \label{VRI}
   \end{figure}

\begin{figure}
   \centering
   \includegraphics[angle=0,width=9.3cm]{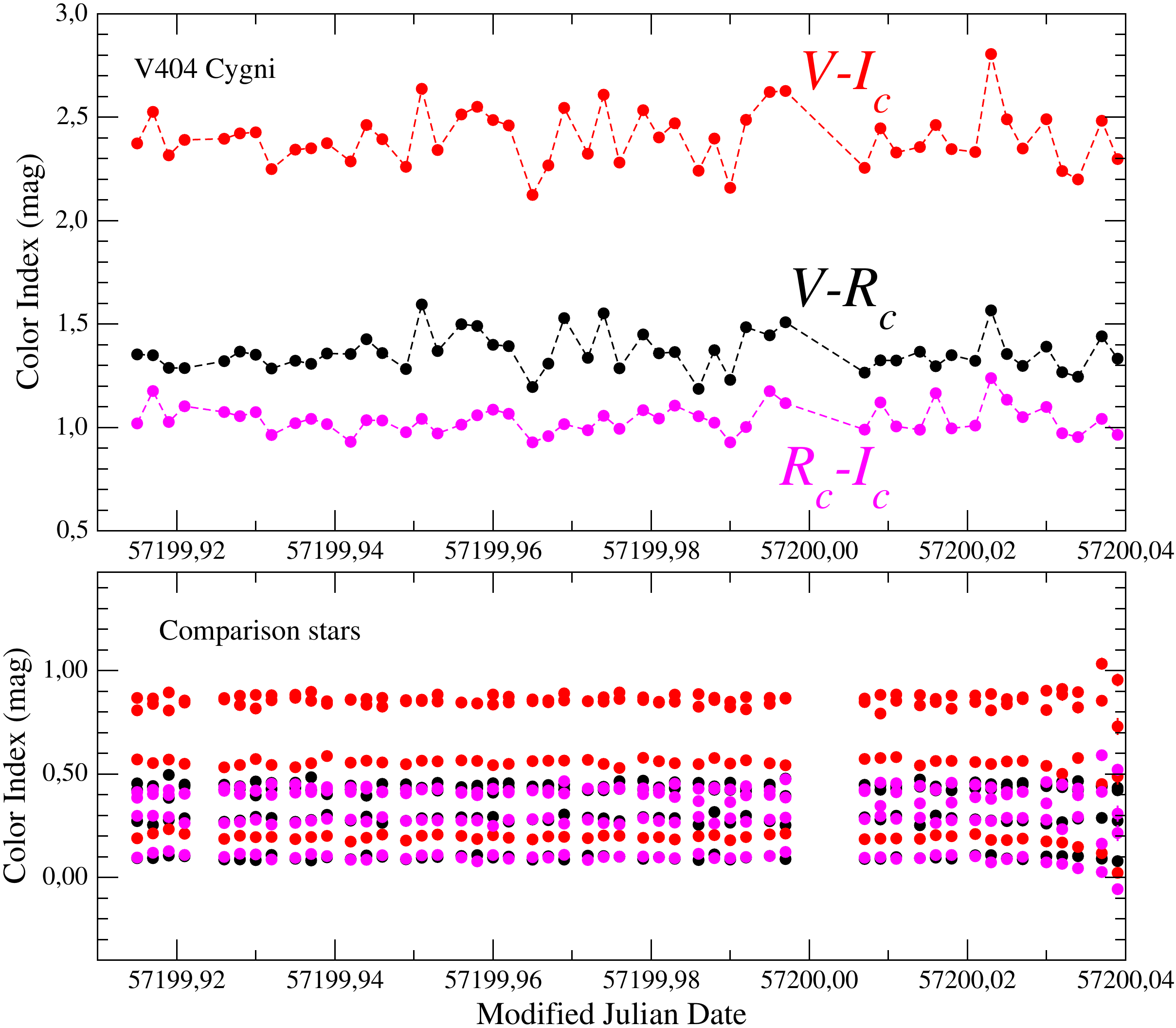}
      \caption{{\bf Top.} Variability of colour indices $V-R$, $V-I$ and $R-I$ of \object{V404 Cygni} in outburst as observed with the UJT on 26 June 2015.
      They are shown plotted in black, red, and magenta colours, respectively. 
  {\bf Bottom.} The same kind of plot for the four comparison stars plotted at the same scale.
  Their colours remained constant within 0.02-0.03 mag.   
    }
         \label{colours_VRI}
   \end{figure}

Differential $VR_cI_c$ photometry was performed on \object{V404 Cygni} during 3 h
with exposure times of 60 s in each filter. The total number of measurements acquired per filter was $N_{\rm obs} = 51$. Four 
comparison stars in the field were used
whose photometric behaviour, within 0.01-0.02 mag,  was found to be very stable in all bands. Their 
$VR_c I_c$ magnitudes were retrieved from the AAVSO database\footnote{American Association of Variable Stars Observers, http://www.aavso.org
 and references therein.},
and are given in Online Table \ref{comparisons}. Photometry in $U$ and $B$ bands was not acquired because of very low source counts.
For each observed photometric band, here generically indicated by an effective wavelength $\lambda$, the magnitudes of the variable target $m^{\rm var}_{\lambda}$
 were derived with respect to the comparison stars according to:
\begin{equation}
m^{\rm var}_{\lambda} = m^{\rm com}_{\lambda} + \Delta m_{\lambda, {\rm ins}} + T_{\lambda, C_{\lambda}} \Delta C_{\lambda, {\rm ins}}, \\  \label{trans}
\end{equation}
where $ m^{\rm com}_{\lambda}$ is the  magnitude of the comparison star being used, and
$\Delta m_{\lambda, {\rm ins}}$ 
and
$\Delta C_{\lambda, {\rm ins}}$ 
are the instrumental differences in magnitude and colour between the target and comparison star, respectively. 
The factor $T_{\lambda, C_{\lambda}}$ is the colour transformation coefficient that was separately determined using standards in clusters and other fields.
The colour $(V-R_c)$ was used in Eq. \ref{trans} for $V$- and $R_c$-band observations, while the colour $(V-I_c)$ was preferred for $I_c$-band observations instead.
The system photometric properties are approximately stable from night to night and we typically obtain $|T_{\lambda, C_{\lambda}}| <  0.1$.
Finally, the different measurements of $m^{\rm var}_{\lambda}$ were weighted according to their respective uncertainty and then averaged. The corresponding results are
presented in Online Table \ref{data} and Figs. \ref{VRI} and \ref{colours_VRI}. The errors quoted here are representative of the differential photometry process, and do not include the uncertainty in the
absolute calibration of the comparison stars ($\sim 0.02$ mag).
Main flaring events in Fig. \ref{VRI} are labelled from 1 to 6 for later discussion.

\onltab{

\begin{table*}
\caption{AAVSO comparison stars used in this work.}    
\label{comparisons}      
\centering                        
\begin{tabular}{cccccccc}     
\hline\hline             
AAVSO   & R.A.(J2000.0) & DEC.(J2000.0) &  $B$   &  $ V$   &  $R_c$ & $I_c$  \\
     Id.       &  (hms)        &   (dms)       & (mag)  &  (mag)   &  (mag)   & (mag)  \\  
\hline
125 &  20:23:56.47 & +33:48:16.9  & 12.976  0.019  &  12.497  0.014  &   12.204  0.018 & 11.926  0.018  \\
128 &  20:24:07.24 & +33:50:52.2  & 13.499  0.017  &  12.815  0.014  &   12.384  0.018 & 11.967  0.018  \\
132 &  20:24:08.89 & +33:54:38.6  & 13.879  0.018  &  13.164  0.012  &   12.712  0.016 & 12.284  0.016  \\
134 &  20:23:53.43 & +33:52:24.6  & 14.467  0.017  &  13.361  0.015  &   12.737  0.019 & 12.203  0.019  \\
   \hline
\end{tabular}
 \end{table*}


\begin{table*}
\caption{Optical photometry of \object{V404 Cygni} wiht the UJT telescope.}          
\label{data}      
\centering                        
\begin{tabular}{ccccccccc}     
\hline\hline             
MJD$^{*}$        & $V$-band     &    MJD$^{*}$  & $R_c$-band    &   MJD$^{*}$    &  $I_c$-band  \\
           & (mag)        &         & (mag)       &          &  (mag)    \\
\hline
 57199.91379  & $12.842  \pm  0.012 $ & 57199.91456 & $11.488  \pm 0.007 $ & 57199.91532 &  $10.468  \pm  0.009$ \\
 57199.91610  & $12.825  \pm  0.013 $ & 57199.91686 & $11.476  \pm 0.008 $ & 57199.91763 &  $10.299  \pm  0.010$ \\
 57199.91840  & $12.547  \pm  0.011 $ & 57199.91916 & $11.258  \pm 0.008 $ & 57199.91993 &  $10.231  \pm  0.012$ \\
 57199.92070  & $12.613  \pm  0.013 $ & 57199.92146 & $11.325  \pm 0.007 $ & 57199.92223 &  $10.222  \pm  0.009$ \\
 57199.92478  & $12.797  \pm  0.014 $ & 57199.92554 & $11.476  \pm 0.006 $ & 57199.92631 &  $10.401  \pm  0.009$ \\
 57199.92708  & $12.731  \pm  0.015 $ & 57199.92785 & $11.364  \pm 0.006 $ & 57199.92861 &  $10.309  \pm  0.010$ \\
 57199.92938  & $12.707  \pm  0.013 $ & 57199.93015 & $11.354  \pm 0.006 $ & 57199.93091 &  $10.280  \pm  0.009$ \\
 57199.93169  & $12.688  \pm  0.012 $ & 57199.93245 & $11.402  \pm 0.010 $ & 57199.93322 &  $10.438  \pm  0.010$ \\
 57199.93399  & $12.912  \pm  0.012 $ & 57199.93476 & $11.589  \pm 0.008 $ & 57199.93552 &  $10.568  \pm  0.012$ \\
 57199.93629  & $12.948  \pm  0.016 $ & 57199.93706 & $11.640  \pm 0.006 $ & 57199.93782 &  $10.598  \pm  0.011$ \\
 57199.93860  & $13.013  \pm  0.018 $ & 57199.93936 & $11.655  \pm 0.007 $ & 57199.94013 &  $10.639  \pm  0.010$ \\
 57199.94090  & $13.095  \pm  0.013 $ & 57199.94167 & $11.739  \pm 0.009 $ & 57199.94243 &  $10.808  \pm  0.009$ \\
 57199.94320  & $13.241  \pm  0.016 $ & 57199.94397 & $11.814  \pm 0.008 $ & 57199.94473 &  $10.779  \pm  0.009$ \\
 57199.94551  & $13.241  \pm  0.017 $ & 57199.94627 & $11.881  \pm 0.008 $ & 57199.94704 &  $10.847  \pm  0.011$ \\
 57199.94781  & $13.455  \pm  0.014 $ & 57199.94858 & $12.172  \pm 0.010 $ & 57199.94934 &  $11.194  \pm  0.010$ \\
 57199.95012  & $13.539  \pm  0.018 $ & 57199.95088 & $11.944  \pm 0.008 $ & 57199.95165 &  $10.901  \pm  0.012$ \\
 57199.95252  & $13.401  \pm  0.018 $ & 57199.95328 & $12.031  \pm 0.007 $ & 57199.95405 &  $11.060  \pm  0.012$ \\
 57199.95482  & $13.480  \pm  0.018 $ & 57199.95559 & $11.980  \pm 0.009 $ & 57199.95635 &  $10.966  \pm  0.009$ \\
 57199.95712  & $13.391  \pm  0.013 $ & 57199.95789 & $11.900  \pm 0.008 $ & 57199.95866 &  $10.841  \pm  0.010$ \\
 57199.95943  & $13.194  \pm  0.014 $ & 57199.96019 & $11.794  \pm 0.007 $ & 57199.96096 &  $10.708  \pm  0.011$ \\
 57199.96173  & $12.977  \pm  0.011 $ & 57199.96250 & $11.584  \pm 0.008 $ & 57199.96326 &  $10.517  \pm  0.010$ \\
 57199.96404  & $12.920  \pm  0.011 $ & 57199.96480 & $11.724  \pm 0.007 $ & 57199.96557 &  $10.796  \pm  0.010$ \\
 57199.96634  & $13.333  \pm  0.013 $ & 57199.96711 & $12.024  \pm 0.006 $ & 57199.96787 &  $11.065  \pm  0.009$ \\
 57199.96865  & $13.631  \pm  0.016 $ & 57199.96941 & $12.102  \pm 0.006 $ & 57199.97018 &  $11.086  \pm  0.011$ \\
 57199.97095  & $13.608  \pm  0.020 $ & 57199.97172 & $12.271  \pm 0.007 $ & 57199.97248 &  $11.285  \pm  0.015$ \\
 57199.97326  & $13.677  \pm  0.021 $ & 57199.97402 & $12.126  \pm 0.007 $ & 57199.97479 &  $11.068  \pm  0.012$ \\
 57199.97556  & $13.444  \pm  0.014 $ & 57199.97632 & $12.157  \pm 0.007 $ & 57199.97709 &  $11.163  \pm  0.011$ \\
 57199.97786  & $13.500  \pm  0.016 $ & 57199.97863 & $12.050  \pm 0.006 $ & 57199.97939 &  $10.966  \pm  0.009$ \\
 57199.98017  & $13.354  \pm  0.013 $ & 57199.98093 & $11.995  \pm 0.007 $ & 57199.98170 &  $10.952  \pm  0.012$ \\
 57199.98247  & $13.166  \pm  0.013 $ & 57199.98324 & $11.801  \pm 0.007 $ & 57199.98400 &  $10.695  \pm  0.011$ \\
 57199.98478  & $13.059  \pm  0.014 $ & 57199.98554 & $11.872  \pm 0.007 $ & 57199.98631 &  $10.818  \pm  0.009$ \\
 57199.98708  & $13.235  \pm  0.011 $ & 57199.98785 & $11.862  \pm 0.007 $ & 57199.98861 &  $10.838  \pm  0.010$ \\
 57199.98938  & $13.383  \pm  0.018 $ & 57199.99015 & $12.152  \pm 0.006 $ & 57199.99091 &  $11.224  \pm  0.010$ \\
 57199.99169  & $13.720  \pm  0.014 $ & 57199.99245 & $12.235  \pm 0.007 $ & 57199.99322 &  $11.233  \pm  0.013$ \\
 57199.99399  & $13.696  \pm  0.014 $ & 57199.99476 & $12.250  \pm 0.007 $ & 57199.99552 &  $11.074  \pm  0.014$ \\
 57199.99629  & $13.354  \pm  0.012 $ & 57199.99706 & $11.845  \pm 0.007 $ & 57199.99782 &  $10.727  \pm  0.011$ \\
 57200.00593  & $13.287  \pm  0.018 $ & 57200.00669 & $12.022  \pm 0.007 $ & 57200.00746 &  $11.032  \pm  0.012$ \\
 57200.00823  & $13.317  \pm  0.012 $ & 57200.00900 & $11.992  \pm 0.006 $ & 57200.00976 &  $10.870  \pm  0.011$ \\
 57200.01054  & $13.285  \pm  0.011 $ & 57200.01130 & $11.961  \pm 0.006 $ & 57200.01207 &  $10.955  \pm  0.012$ \\
 57200.01284  & $13.297  \pm  0.013 $ & 57200.01361 & $11.931  \pm 0.008 $ & 57200.01437 &  $10.941  \pm  0.012$ \\
 57200.01514  & $13.457  \pm  0.012 $ & 57200.01591 & $12.160  \pm 0.007 $ & 57200.01667 &  $10.994  \pm  0.011$ \\
 57200.01745  & $13.291  \pm  0.013 $ & 57200.01821 & $11.941  \pm 0.006 $ & 57200.01898 &  $10.945  \pm  0.013$ \\
 57200.01975  & $13.588  \pm  0.016 $ & 57200.02052 & $12.265  \pm 0.007 $ & 57200.02128 &  $11.256  \pm  0.013$ \\
 57200.02205  & $13.550  \pm  0.021 $ & 57200.02282 & $11.984  \pm 0.007 $ & 57200.02358 &  $10.745  \pm  0.014$ \\
 57200.02435  & $12.914  \pm  0.014 $ & 57200.02512 & $11.559  \pm 0.008 $ & 57200.02588 &  $10.425  \pm  0.012$ \\
 57200.02665  & $12.755  \pm  0.009 $ & 57200.02742 & $11.457  \pm 0.007 $ & 57200.02818 &  $10.406  \pm  0.013$ \\
 57200.02896  & $12.827  \pm  0.014 $ & 57200.02972 & $11.436  \pm 0.006 $ & 57200.03049 &  $10.337  \pm  0.014$ \\
 57200.03126  & $12.646  \pm  0.016 $ & 57200.03203 & $11.378  \pm 0.006 $ & 57200.03279 &  $10.406  \pm  0.016$ \\
 57200.03357  & $12.986  \pm  0.011 $ & 57200.03433 & $11.741  \pm 0.010 $ & 57200.03510 &  $10.787  \pm  0.016$ \\
 57200.03587  & $13.294  \pm  0.015 $ & 57200.03663 & $11.853  \pm 0.008 $ & 57200.03740 &  $10.811  \pm  0.016$ \\
 57200.03817  & $13.249  \pm  0.015 $ & 57200.03894 & $11.916  \pm 0.008 $ & 57200.03970 &  $10.951  \pm  0.021$ \\
   \hline
\end{tabular}
~\\
($^*$) The Modified Julian Date (MJD) given is heliocentric and corresponds to the mid-exposure time.
 \end{table*}


%
\begin{table*}
\caption{Recovered artificial time lags with interpolated sampling CCF for $V$-band data.}          
\label{artificial_lag}      
\centering                        
\begin{tabular}{cccc}     
\hline\hline                
Artificial time lag  & Recovered time lag & Artificial time lag & Recovered time lag  \\  
(s)    &   (s)  \\
\hline          
0    	& $-0.02$   &    60	& 52.92                \\
1    	& $-0.02$   &    70	& 62.85                \\
5       & $-0.02$   &    80	& 72.80                \\
10  	& 3.22      &    90	& 82.73                \\
20  	& 13.15     &    100    & 92.66                \\
30  	& 23.08     &    110	& 102.59                \\
40      & 33.06     &    120	& 112.52               \\
50	& 42.99     &    1000	&  971.77              \\
  \hline
 \hline                                  
\end{tabular}

\end{table*}
}

\section{Discussion}

The light curves in Fig. \ref{VRI} and \ref{colours_VRI} clearly show how \object{V404 Cygni} varied and flared intensively
during the observation. The amplitude of variation was as large as one magnitude and several tenths of a magnitude
 in brightness and colour, respectively. These variations occurred 
on timescales as short as 10 minutes. 
The significance and intrinsic origin
of these flares is ensured given that the comparison stars
remained practically flat within a few hundreds of a magnitude.
This agrees well with the behaviour reported by different observers at other times
during the intensive coverage of the outburst   \citep[e.g.][]{2015ATel.7677....1H}.
From causality arguments, our data puts a coarse upper limit of $\sim 1$ AU to the region from where the optical emission arises in the vicinity
of the accretion disk around the \object{V404 Cygni} black hole.

\begin{figure}
   \centering
   \includegraphics[angle=0,width=8.3cm]{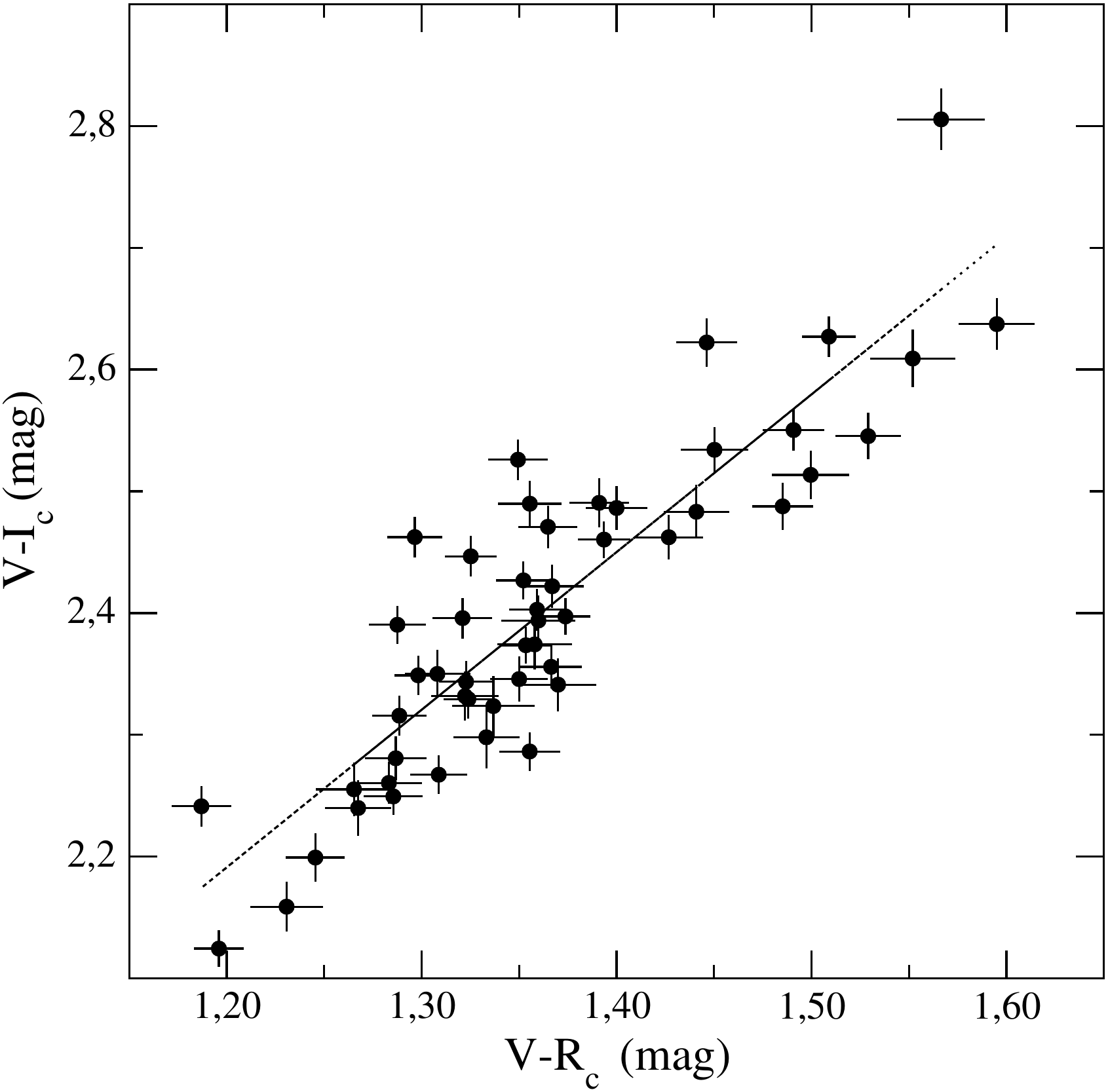}
      \caption{Evolution of \object{V404 Cygni} in the colour-colour plane during a few hours on 26 June 2015 as observer with the UJT.
      The dotted line is a linear regression fit guiding the eye to better appreciate the colour-colour correlation suggested in the text.
    }
         \label{colourevol}
   \end{figure}

It is also very interesting that
the observed variability apparently followed a clear pattern in colour-colour diagrams.
This is illustrated in Fig. \ref{colourevol} where the pattern becomes more evident as the source evolved in a very restricted
region of the diagram. A simple regression fit yields a correlation coefficient as high as 0.89, thus suggesting that a strong connection between the source colours existed during the flaring events.
The modelling of such behaviour is beyond the observational scope of this paper, and we limit ourselves here
to state this observational fact. Nevertheless, we speculate that it could be intimately linked to
the source spectral evolution discussed below.

\begin{figure}
   \centering
   \includegraphics[angle=0,width=9.0cm]{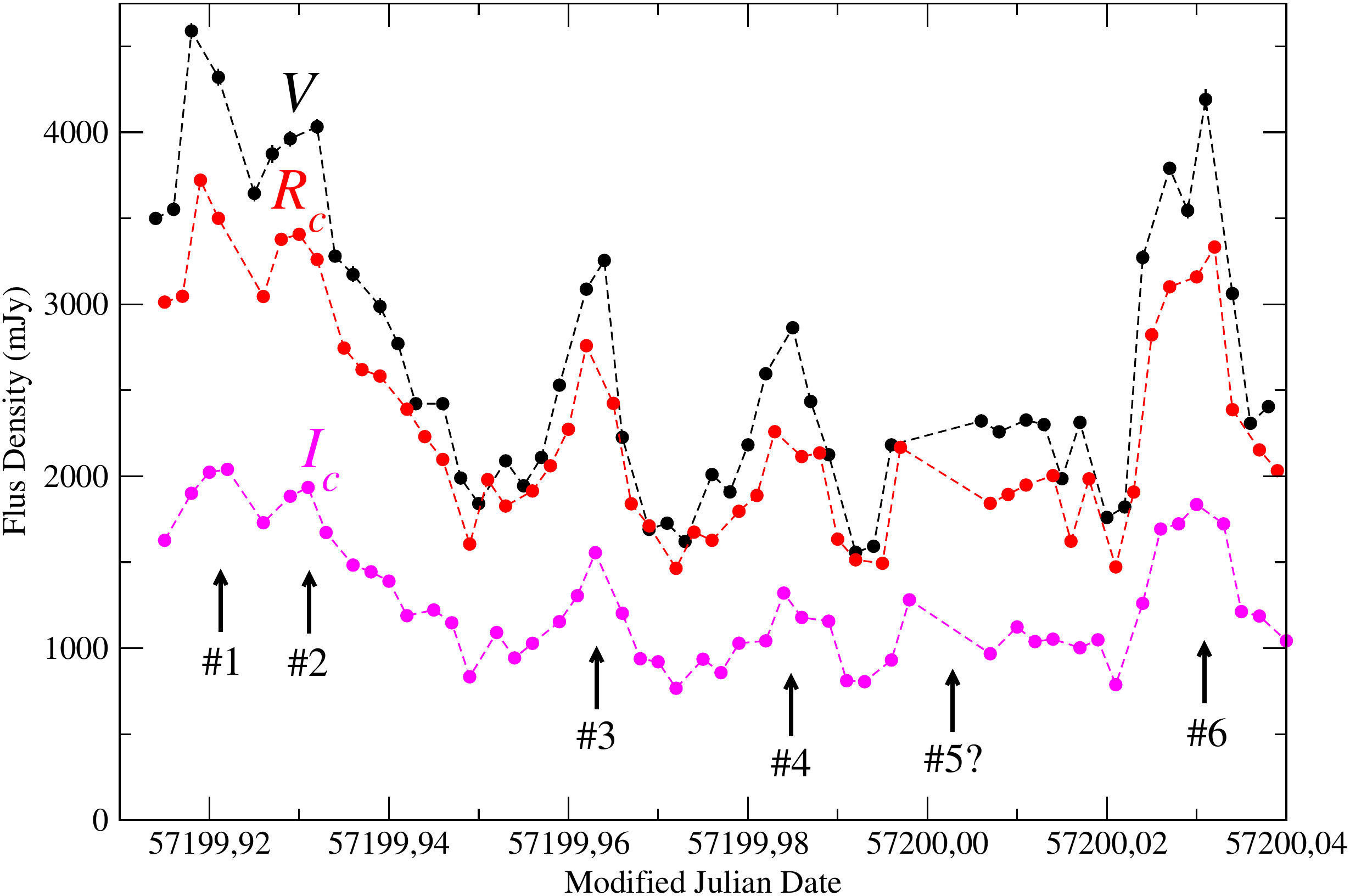}
      \caption{Dereddened light curves of \object{V404 Cygni} in outburst as observed with the UJT on 26 June 2015  in the $V$, $R_c$ and $I_c$ bands
      assuming an interstellar extinction of $A_V=4.0$ mag. The brightness level has been converted from magnitudes to a flux density scale, in mJy units.
      Flares are labelled as in Fig. \ref{VRI}.
    }
         \label{VRIdered}
   \end{figure}

The intrinsic amplitudes of the observed flares become more obvious when de-reddening the photometric observations, 
and expressing them in terms of flux density. The interstellar extinction law in  \citet{2000asqu.book.....C} was used for de-reddening purposes.
This is presented in Fig. \ref{VRIdered}, where two isolated flares (\#3 and \#4) are easily visible with their peaks separated by nearly a half hour. 
Other flaring events  both at the beginning  (\#1 and \#2) and the end (\#5 and \#6) of the observation are also present
within comparable time intervals. However, their individual evolution
cannot be so easily disentangled because they are close in time or not well sampled.
By analogy with other similar flaring behaviours,
 such as in the LMXB \object{GRS 1915+105} at radio and near-infrared wavelengths \citep{1998A&A...330L...9M},
one is tempted to interpret
this recursive flare pattern as episodes of replenishment and
emptying of the inner accretion disk. These kind of events 
come accompanied by the ejection of plasmons along collimated jets as a result of accretion disk instabilities.
These plasma clouds typically have a
characteristic non-thermal, synchrotron
emission mechanism. Their spectrum spans from radio to much shorter wavelengths depending on the energy cut-off of relativistic electrons and the strength of magnetic field.
The fact that strong radio emission from \object{V404 Cygni} was detected  in 1.4 GHz observations carried out just six hours before ours, peaking at $0.364 \pm 0.030$ Jy
\citep{2015ATel.7733....1T}, would be naturally understood in this context.
The similarity between \object{V404 Cygni} and \object{GRS 1915+105} has already been noted by \citet{2015ATel.7658....1M}.

The observed optical flares appeared superimposed onto an average pedestal emission level likely due to hot 
 accretion disk in \object{V404 Cygni}. From Fig. \ref{VRIdered}, the 
average de-reddened flux densities were found to be
$2.27 \pm 0.16$, $2.02 \pm 0.11$, and $1.10\pm 0.04$ Jy in the $V$, $R_c$ and $I_c$ bands, respectively. By fitting a simple power law,
the average spectral energy distribution (SED) of \object{V404 Cygni} 
depends on frequency roughly as $\propto \nu^{2.1\pm0.3}$. No correction for the contribution of the H$\alpha$ emission line
to $R_c$-band data has been attempted here.
The resulting power law index is consistent with the UJT data sampling the Raleigh Jeans $\nu^2$ part of the accretion disk spectrum.
Extrapolation of this power law to the radio band falls orders of magnitude below the reported radio flux densities. 
Therefore, an additional emission component must exist to explain the \citet{2015ATel.7733....1T} radio detection.
As stated before, this component is most naturally interpreted as originating in plasma ejection events. 

In the scenario outlined here, one would expect the peak time of an individual flare to be dependent on the wavelength of observation.
According to the simple \citet{1966Natur.211.1131L} model of a synchrotron-emitting expanding plasmon, the light curve observed
at a  wavelength $\lambda$ reaches its maximum after a time given by
\begin{equation}
t_{m,\lambda} = t_{m,\lambda_0} [\lambda / \lambda_0]^{-\frac{p+4}{4p+6}},  \label{delays}
\end{equation}
where $t_{m,\lambda_0}$ is the time of maximum at a reference wavelength $\lambda_0$, and $p$ is the power-law index of the energy distribution
of relativistic electrons. 

In the absence of contemporaneous high time resolution radio photometry, where a $\lambda$-dependent delay  would be easier to measure, 
we searched our optical data for possible time lags between the different light curves at $V$, $R_c$ and $I_c$ bands.
Taking, for instance, the isolated flare \#3, 
which peaks around MJD 57199.965 in Figs. \ref{VRI} and \ref{VRIdered}, the rising time from the pedestal emission level
to the flare peak in the $V$-band is $t_{m,V} \simeq 0.01$ d. 
Assuming $p\simeq 1.0$, and given that the central effective wavelengths of our photometric filer set 
($\lambda_V =0.545$ $\mu$m, $\lambda_{R_c}=0.641$ $\mu$m, and $\lambda_{I_c}=0.798$ $\mu$m), we can estimate the expected time delays
in the UJT observations. Using Eq. \ref{delays}, and taking the $V$ band as reference, the predicted delays are 1.2 min and 3.0 min for the $R_c$ and $I_c$ band, respectively.
Similarly, a 1.8 min delay would be expected between $I_c$ and $R_c$ data. Assuming other reasonable values of $p$ 
does not change significantly
these numbers. For instance, taking $p=2.5$ one gets delays of 1.0 min, 2.4 min, and 1.4 min, respectively.

 What happens with real data? 
 To answer this question, we performed a  cross-correlation function (CCF) exercise between the different UJT light curves.
 A problem arises here because the data are not evenly spaced in time.
 The CCF calculation  was then  carried out in two different ways. The first calculation was specially designed for such irregular time series and is based on averaging
data products with similar time lags within a user defined binning \citep{1988ApJ...333..646E}. The second calculation simply makes a linear
interpolation to obtain a regular sampling of the data from which the traditional estimator of the CCF is computed. In our case, this second
approach is justified because our observations were nearly evenly spaced.
When applying the first method, the resulting CCFs do not have enough resolution to clearly establish a non-zero time lag although hints 
of delay between the longer and shorter wavelength filters were obtained.
In contrast, the interpolation method provided well-defined CCF maxima shifted from zero (Online Figs. \ref{correlos_a} and \ref{correlos_b}).
Here, the key fact is that all maxima occur with a clearly negative lag, i.e. consistent with the longer wavelength light curves
that are delayed with respect to the shorter wavelength curves, as expected. Interestingly, similar negative time lags have also been observed in some of the \object{V404 Cygni} 
flares seen by the {\it INTEGRAL} satellite \citep{2015A&A...581L...9R}, where the optical $V$-band emission was delayed with respect  to hard X-rays and soft $\gamma$-rays
from 1.5 to 20-30 min. However, these lags correspond to extremely different energy bands in contrast to our
results here reported within the optical domain.


The measured offset of the maxima with respect to the origin  (in Online Figs. \ref{correlos_a} and \ref{correlos_b})
indicates that the $I_c$ band light curve is delayed with respect to $V$ band 
by $108 \pm 1$ s, while the $R_c$ band lags it by $34\pm1$ s.  
Concerning the two reddest filters, $I_c$ band also lags the $R_c$ band by $34\pm1$ s.
The uncertainties in the estimates of the CCF peak positions were derived using the \citet{1994PASP..106..879W} formula
\begin{equation}
\Delta \tau = \frac{0.38 W_{\rm CCF}}{1+r_{\rm max} (N_{\rm total}-2)^{1/2}},  \label{ccf_error}
\end{equation}
where $W_{\rm CCF}$ is the full width at half maximum of the CCF peak with $r_{\rm max}$ amplitude, and $N_{\rm total}= (N_{\rm obs}-1) N $ the full number
of available points, where $N$ is the amount of sampling points.
 In our case, $W_{\rm CCF} \simeq 1000$ s, $r_{\rm max}\simeq 1$, 
and we interpolated up to $N=10^4$ points.
 
Although these numbers are not identical to the values anti\-cipated using Eq. \ref{delays}, 
all of them appear to be in the expected sense and with the expected order of magnitude.
Such agreement is remarkable given the simplicity of the Van der Laan model and the limitation of our data.

In order to assess the robustness of our finding, we performed a sensitivity analysis of how interpolated sampling affects the results.
As seen in Online Fig. \ref{correlos_c}, the derived lags always reach a constant level after a sufficiently large number of interpolated points.
Computing the rms dispersion of the latest 1000 points provides values of  $\sim1$ s comparable to the uncertainty estimates
derived with Eq. \ref{ccf_error}. Online Fig. \ref{correlos_c} also shows that, even with coarse interpolation, the $V$ vs $I_c$ time lag  still 
remains significant although at a lower level. 
The $V$ vs $I_c$ time lag amounts to $86 \pm 16$ s for $N=10$ interpolated sampling points. The other filter combinations,
 $V$ vs $R_c$ and $R$ vs $I_c$, which are closer in wavelength,  remain consistent with zero both amounting to $22\pm 16$ s.

Moreover, to further ensure the confidence on the lag estimates with interpolated light curves,
we conducted a series of simulations by cross-correlating the observed light curves with themselves
after introducing different artificial time lags (see Online Table \ref{artificial_lag}).
 Our procedure was able to confidently recover these time lags, within a few seconds, but always with a systematic negative bias that
 results as an intrinsic effect of the interpolation method used. In our case this is not expected to introduce 
 a severe effect because of the lack of skewness in the time sampling distribution \citep{npg-18-389-2011}.
 Therefore, the measured lag values quoted
 above should be actually interpreted as lower limits.
  These two consistency checks reinforce our
confidence in the lag detection provided that interpolation is actually a good approximation.

Another test that could be performed is the dependence of the maximum flux densities $S_{m, \lambda}$ as a function of wavelength.
Following \citet{1966Natur.211.1131L}, the expected dependence is:
\begin{equation}
S_{m,\lambda} = S_{m,\lambda_0} [\lambda / \lambda_0]^{-\frac{7p+3}{4p+6}},  \label{ratios}
\end{equation}
Considering the same isolated flare as above, and removing the pedestal flux density with a simple linear fit, we estimate the 
$V$-band maximum
as $S_{m,V} \simeq 1480$ mJy. For $p=1$, one would then predict $S_{m,R_c} \simeq 1260$ mJy  and $S_{m,I_c} \simeq 1010$ mJy.
Our $R_c$ and $I_c$ light curves in Fig. \ref{VRIdered} were not well sampled to appropriately catch their maxima.  With the pedestal flux density removed they
provide $S_{m,R_c}$\gtsima$1020$ mJy and $S_{m,V}$\gtsima$650$ mJy, respectively. While the expected values are not exactly reproduced,
the results are still compatible.

All together, we consider that this finding is supportive of
 our tentative interpretation of the optical flares
as due to non-thermal plasmon ejection events. 
Similar flares in the prototypical case of \object{GRS 1915+105} could not be observed in the optical
 because of the higher interstellar extinction.


\onlfig{

 \begin{figure}
   \centering
   \includegraphics[angle=0,width=10.0cm]{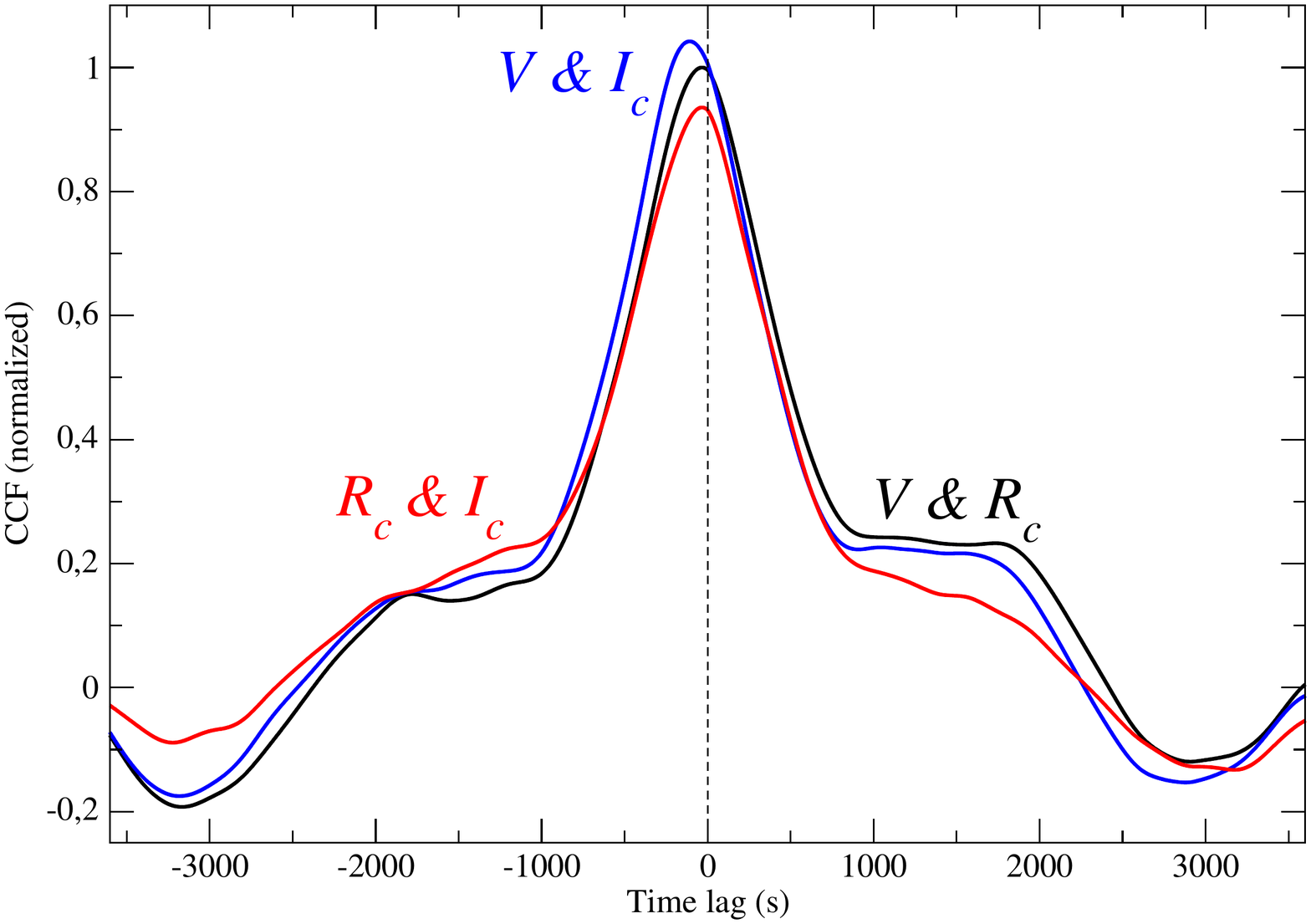}
      \caption{
      CCFs  of the optical light curves $V$ vs $R_c$ (black), $V$ vs $I_c$ (blue) and $R_c$ vs $I_c$ (red). 
      The vertical dashed line corresponds to zero time lag.
      The $V$ vs $R_c$ CCF is normalized to unity while the other two have been slightly scaled above and below for easier display.
    }
         \label{correlos_a}
   \end{figure} 

\begin{figure}
   \centering
   \includegraphics[angle=0,width=10.0cm]{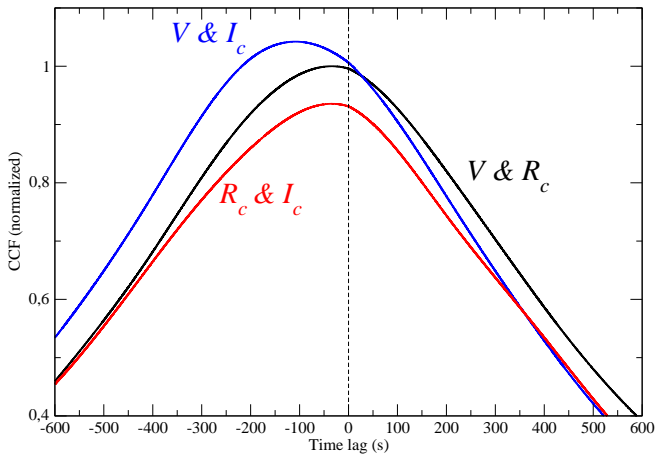}
      \caption{
   Zoomed view of the central maxima of the CCFs  where a clear asymmetry is visible. 
   Colours and normalization are as in Online Fig. \ref{correlos_a}.
    }
         \label{correlos_b}
   \end{figure} 

\begin{figure}
   \centering
   \includegraphics[angle=0,width=10.0cm]{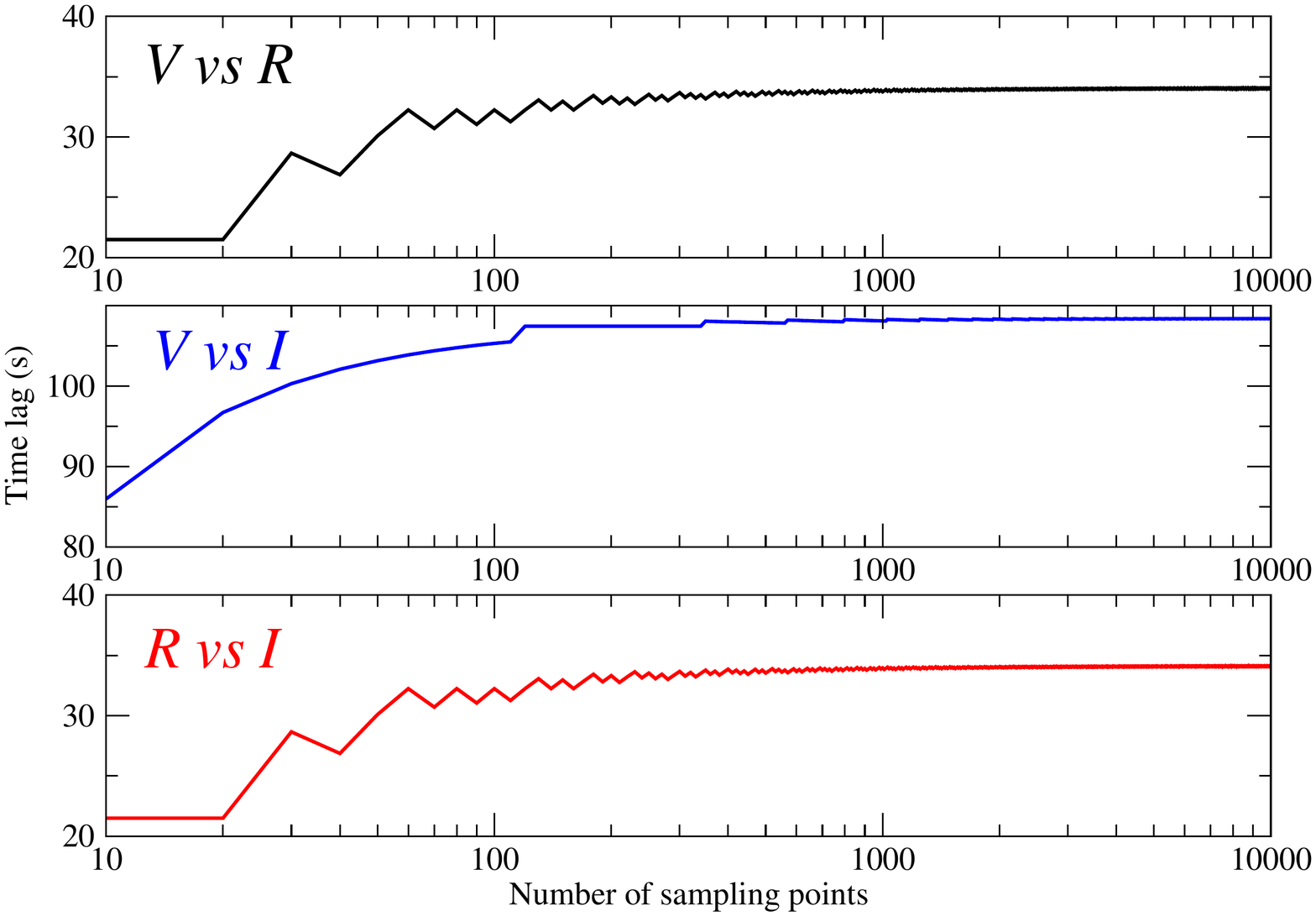}
      \caption{
      Dependence of the derived time lags using the cross-correlation technique on the amount of interpolated sampling of the data. 
      Colours   are as in Figs. \ref{correlos_a} and \ref{correlos_b}.
    }
         \label{correlos_c}
   \end{figure} 

  }


In addition, the CCFs in Online Fig. \ref{correlos_a} display clear secondary maxima
corresponding to time lags of $\sim1500$ s. This agrees with the idea, already expressed before, that the optical light curve of \object{V404 Cygni} 
in outburst during our observation consists of the
superposition of  flares that appear every half hour on average. 
Flaring behaviour with a similar timescale is also present in
optical light curves reported by other observers just  before and after
our UJT observations  \citep{2015ATel.7721....1S, 2015ATel.7737....1S}.

The energy involved in each ejection event can also be derived based on simple equipartition arguments using the \citet{pacholczyk1970radio} formulation.
Looking at Fig. \ref{VRIdered}, 
 the  flare incremental flux density  with respect to the pedestal value is in the range $\sim 0.5$-1.0 Jy, and this occurs typically 
 $\sim 10^3$ s after the estimated onset of the flares.
Such flux density increments are comparable to the 1.4 GHz emission levels  reported near the time of
our observations within a factor of a few. This renders the possibility of a nearly flat synchrotron component from radio to optical wavelengths conceivable.
Assuming an expansion velocity of $\sim 0.5c$, the plasmon size at the time of maximum would be $\sim 1.5 \times 10^{13}$ cm.
This is  equivalent to 0.4 milli-arcsec
at the \object{V404 Cygni} distance and consistent with previous size upper limits. 
Using this dimension, the radio/optical flux densities quoted before, and assuming relativistic electrons with Lorentz
factor $\gamma \leq 10^4$, the total energy content is estimated to be $2\times 10^{40}$ erg, with a magnetic field of 12 G. The source brightness
temperature does not exceed a few  $10^{12}$ K. Under all such assumptions,
the corresponding non-thermal luminosity, from the radio to optical domain, amounts to $2.2 \times 10^{36}$ erg s$^{-1}$, which implies a synchrotron life-time
of about 2 h for individual flaring events. This is only four times their observed recurrence interval, thus flare superposition should become negligible after a few flaring events.
Finally, the plasmon mass involved in the ejection is found to be $\sim 2 \times 10^{20}$ g provided that there is one proton per relativistic electron.
These numbers do not differ by more than one order of magnitude when compared with the reference case of \object{GRS 1915+105}.


\section{Conclusions}

We have reported new optical photometric data for the LMXB \object{V404 Cygni} during its 2015 June  outburst obtained with a small educational telescope. The time interval
covered by our observations complements and contributes to the multi-wavelength campaign carried out  as a world-wide effort by many observers.
From the analysis of our data alone, some interesting findings can already be advanced and we summarize them as follows.


The UJT optical light curves  consisted of consecutive and partially overlapping  flaring events,  appearing about every half hour and
with very large amplitudes  ($\sim 1$ mag).
Timescales of  variability as short as 10 minutes could be clearly detected both in brightness and colour.
The optical variability observed appears to display a correlation pattern in the colour-colour plane, 
whose interpretation will deserve future
theoretical work.

The de-reddened optical continuum had an average positive spectral index close the $+2$ value. 
Thus, for  most of the time
the observed photometric bands  ($VR_cI_c$) were mainly sampling the Rayleigh Jeans part of the accretion disk spectrum.
The \object{V404 Cygni} flares, superposed on this continuum, are tentatively interpreted as non-thermal flaring events.
With all caution, we suggest that the flares are due to 
relativistic plasmons ejected following successive replenishing and emptying episodes of the inner accretion disk.
The fact that non-thermal, synchrotron emission appears to contribute to optical wavelengths renders \object{V404 Cygni} a very interesting system
to better study the energetics of black hole LMXBs.


From a CCF analysis, we have found a systematic time lag between the UJT optical light curves acquired with the different  filters during our observations.
These lags are such that emission at  shorter wavelengths precedes emission at longer wavelengths by $\sim 1$ min.
This finding was most evident when light curves were resampled and evenly interpolated, which we believe is a reasonable approach 
when dealing with nearly evenly spaced data.
This delay is in qualitative agreement with a simple Van der Laan model for the expansion of a synchrotron emitting plasmon,  
and the relative amplitude of flare maxima is also consistent.
An estimate of the plasmon physical parameters required for the non-thermal spectrum to extend from radio to optical wavelengths 
appears to be similar to other well-known systems, such as \object{GRS 1915+105}.



\begin{acknowledgements}
    This work was supported by grant AYA2013-47447-C3-3-P from the Spanish Ministerio de Econom\'{\i}a y Competitividad (MINECO),
and by the Consejer\'{\i}a de Econom\'{\i}a, Innovaci\'on, Ciencia y Empleo of Junta de Andaluc\'{\i}a
under research group FQM-322, as well as FEDER funds.  
    
\end{acknowledgements}

%
%



\bibliographystyle{aa} 
\bibliography{references} 
     

\end{document}